# Gradings, Braidings, Representations, Paraparticles: Some Open Problems

**Konstantinos Kanakoglou**

School of Mathematics, Aristotle University of Thessaloniki (AUTH), Thessaloniki 54124, Greece;
e-mail: kanakoglou@hotmail.com, kanakoglou@ifm.umich.mx

**Abstract:** A research proposal on the algebraic structure, the representations and the possible applications of paraparticle algebras is structured in three modules: The first part stems from an attempt to classify the inequivalent gradings and braided group structures present in the various parastatistical algebraic models. The second part of the proposal aims at refining and utilizing a previously published methodology for the study of the Fock-like representations of the parabosonic algebra, in such a way that it can also be directly applied to the other parastatistics algebras. Finally, in the third part, a couple of Hamiltonians is proposed, and their sutability for modeling the radiation matter interaction via a parastatistical algebraic model is discussed.

## 1. Introduction

The "free" Paraparticle algebras were introduced in the 1950s by Green [1] and Volkov [2,3] as an alternative—to the Canonical Commutation Relations (CCR) and the Canonical Anti-commutation Relations (CAR)—starting point for the free field quantization, but it was soon realized that these algebras also constitute a possible answer to the "Wigner Quantization scheme" [4]. In the decades that followed, numerous papers have appeared dealing with various aspects of their mathematical and physical implications. Nevertheless, few of them could be characterized as genuine advances:

The first important result for these algebras was the classification of their Fock-like representations: In [5] Greenberg and Messiah determined conditions which uniquely specify a class of representations of the "free" parabosonic $P_B$ and the "free" parafermionic $P_F$ algebras. We are going to call these representations Fock-like due to the fact that they are constructed as generalizations of the usual symmetric Fock spaces of the Canonical Commutation relations (CCR) and the antisymmetric Fock spaces of the Canonical Anticommutation Relations (CAR), leading to generalized versions of the Bose-Einstein and the Fermi-Dirac statistics. In [5] it is shown that the parafermionic Fock-like spaces lead us to a direct generalization of the Pauli exclusion principle. The authors further prove that these representations are parametrized by a positive integer $p$ or, equivalently, that they are classified by the positive integers. However they did not construct analytical expressions for the action of the generators on the specified spaces, due to the intractable computational difficulties inserted by the complexity of the (trilinear) relations satisfied by the generators of the algebra. Apart from some special cases (*i.e.*, single degree of freedom algebras or order of the representations $p = 1$) the problem of constructing explicitly the determined representations remained unsolved for more than 50 years. In the same paper [5], the authors introduced a couple of interacting paraparticle algebras mixing parabosonic and parafermionic degrees of freedom: the Relative Parabose Set $P_{BF}$, the Relative Parafermi Set $P_{FB}$ and the straight Commutation and Anticommutation relations, abbreviated SCR and SAR respectively.

The problems of the explicit construction of the Fock-like representations of the above algebras, in the general case of the infinite degrees of freedom, remained unsolved until recently, due mainly to the serious computational difficulties introduced by the number and the nature of the trilinear relations between the generators of these algebras. The solution to these problems was finally given in a series of papers [6–8]: The authors proceeded—utilizing a series of techniques—to the explicit construction, for an arbitrary value of the positive integer $p$ of the above mentioned Fock-like representations for the $P_B$ anf the $P_F$ algebras. Employing techniques of induced representations, combined with the well known Lie super-algebraic structure of $P_B$ [9] and Lie algebraic structure of $P_F$ [10,11], together with elements from the representation theory of the (complex) Lie superalgebra $osp(1/2n)$ and the (complex) Lie algebra $so(2n+1)$, they proceed to construct Gelfand-Zetlin bases and calculate the corresponding matrix elements. However, the general cases of $P_{BF}$, $P_{FB}$, $S_{BF}$ and $S_{FB}$ algebras remain still open (even in the case of the finite degrees of freedom).

Other interesting and important advances in the study of the algebraic properties of the various Paraparticle algebras have been the studies of the various $(G, \vartheta)$-Lie structures present: The Lie algebraic structure of the Parafermionic algebra $P_F$ had already been known since the time of [10,11]. In the 1980s, the pioneering works of Palev [9] established Lie superalgebraic structures for the Parabosonic algebra $P_B$ and the Relative Parafermi Set $P_{FB}$ algebra [12,13] as well. The picture expands even more with recent results on the $(Z_2 \times Z_2)$ - graded $\vartheta$ -colored Lie structure of the Relative Parabose Set $P_{BF}$ algebra [14,15].

## 1.1. Structure of the Paper

The aim of the present paper is to introduce a research proposal, revolving around the above mentioned topics, trying to describe and extend already open problems, generalize previously obtained results and develop new methodological approaches where this might appear feasible. The project is structured in three modules corresponding to: (a) the study and, if possible, the classification of the graded and braided algebraic structures present in the algebras of parastatistics; (b) the study and the attempt to establish explicit construction of representations for these algebras; and finally (c) a proposal for a Hamiltonian written in terms of paraparticle algebra generators, and targeting the description of the radiation–matter interaction.

In Section 2, we start the elucidation by introducing the paraparticle algebras (and their notation), which are going to constitute the central object of study, in terms of generators and relations: The "free" Parabosonic algebra $P_B$, the "free" Parafermionic algebra $P_F$, the Relative Parabose Set algebra $P_{BF}$ and the Relative Parafermi Set algebra $P_{FB}$, the straight Commutation Relations SCR and the straight anticommutation relations SAR. For the sake of completeness, we also review some more or less well known particle algebras of mathematical physics which are directly related to the proposed methods: the Canonical Commutation Relations (CCR), the Canonical Anticommutation Relations (CAR), the symmetric Clifford-Weyl algebra $W_s$, and the antisymmetric Clifford-Weyl algebra $W_{as}$.

In Section 3, previously obtained results on the $\vartheta$ -color, $G$ -graded Lie algebraic structures of various paraparticle algebras are reviewed and an attempt is made to generalize or extend these results. After a conceptual introduction to the modern algebraic treatment of the notions of grading, and color functions, we focus the discussion on the classification of the actions of group algebras on the paraparticle algebras and the classification of the non-trivial quasitriangular structures of these group algebras rather than on the Lie structures of the paraparticle algebras themselves.

In Section 4, a connection is made with previous results by the author, and a "braided" methodology is outlined for the study and the construction of the representations of the paraparticle algebras. The novel thing in the present approach is the exploitation of the gradings and the braidings of the various particle (CCR and CAR) and paraparticle algebras and their interplay, rather than the use of Lie algebraic techniques followed by other authors [6–8]. We also focus on the description of unsolved mathematical problems, whose solution is a necessary step in order for the method to be finalized in a form applicable to all the paraparticle algebras discussed.

In Section 5, a couple of Hamiltonians is proposed and their suitability for the description of the interaction between a monochromatic parabosonic field and a multiple energy-level system is discussed. Mixed Paraparticle algebras are used as spectrum generating algebras and the idea is based on recent results obtained by the author and other authors, relative to the construction of a class of irreducible representations for a mixed paraparticle algebra combining a single parabosonic and a single parafermionic degree of freedom. The reader with the necessary background in physics literature related to the description of the radiation-matter interaction, will easily recognize that we are actually discussing an attempt to develop a paraparticle multiple-level generalization of the Jaynes-Cummings model [16], which has been a celebrated model of Quantum Optics.

In what follows, all vector spaces, algebras and tensor products will be considered over the field of complex numbers $\mathbb{C}$, the prefix "super" will amount to $Z_2$-graded, $G$ will always stand for a finite, Abelian group, unless stated otherwise, and finally, following traditional conventions of physics literature $[x, y] = xy - yx$ will stand for the commutator and $\{x, y\} = xy + yx$ for the anticommutator. Moreover, the term module will be used as identical to representation and whenever formulas from physics enter the text, we use the traditional convention $\hbar = m = \omega = 1$.

## 2. The Algebras, in Terms of Generators and Relations

In the following table, the various particle and paraparticle algebras used and studied in this paper are presented in generators and relations. In what follows: $i, j, k, l, m = 1, 2, \ldots$ and $\xi, \eta, \varepsilon = \pm$.

| Algebras: / Generators and Relations: | CCR | CAR | $W_s$ | $W_{as}$ | $P_B$ | $P_F$ | $P_{BF}$ | $P_{FB}$ | SCR | SAR |
|---|---|---|---|---|---|---|---|---|---|---|
| $\left[b_i^\varepsilon, b_j^\eta\right] = \frac{1}{2}(\eta-\varepsilon)\delta_{ij}I$ | • | | • | • | | | | | | |
| $\left\{f_i^\varepsilon, f_j^\eta\right\} = \frac{1}{2}\lvert\eta-\varepsilon\rvert\delta_{ij}I$ | | • | • | • | | | | | | |
| $\left[b_i^\varepsilon, f_j^\eta\right] = 0$ | | | • | | | | | | • | |
| $\left\{b_i^\varepsilon, f_j^\eta\right\} = 0$ | | | | • | | | | | | • |
| $\left[\{b_i^\xi, b_j^\eta\}, b_k^\varepsilon\right] = (\varepsilon-\eta)\delta_{jk}b_i^\xi + (\varepsilon-\xi)\delta_{ik}b_j^\eta$ | | | | | • | | • | • | • | • |
| $\left[[f_i^\xi, f_j^\eta], f_k^\varepsilon\right] = \frac{1}{2}(\varepsilon-\eta)^2\delta_{jk}f_i^\xi + \frac{1}{2}(\varepsilon-\xi)^2\delta_{ik}f_j^\eta$ | | | | | | • | • | • | • | • |
| $\left[\{b_i^\xi, b_j^\eta\}, f_k^\varepsilon\right] = 0 = \left[[f_i^\xi, f_j^\eta], b_k^\varepsilon\right]$ | | | | | | | • | • | | |
| $\left[\{f_k^\xi, b_l^\eta\}, b_m^\varepsilon\right] = (\varepsilon-\eta)\delta_{lm}f_k^\xi$ <br> $\left\{[b_k^\xi, f_l^\eta], f_m^\varepsilon\right\} = \frac{1}{2}(\varepsilon-\eta)^2\delta_{lm}b_k^\xi$ | | | | | | | • | | | |
| $\left\{[f_k^\xi, b_l^\eta], b_m^\varepsilon\right\} = (\varepsilon-\eta)\delta_{lm}f_k^\xi$ <br> $\left[[b_k^\xi, f_l^\eta], f_m^\varepsilon\right] = \frac{1}{2}(\varepsilon-\eta)^2\delta_{lm}b_k^\xi$ | | | | | | | | • | | |

The CCR algebra consists of the familiar Canonical Commutation Relations of elementary Quantum mechanics and is widely known under the names of boson algebra or Weyl algebra. Similarly, CAR stands for the Canonical Anticommutation Relations or fermion algebra. The study of the properties and the representations of these algebras constitute some of the oldest problems of Mathematical Physics and their origins are dated since the early days of Quantum theory.

The algebra $W_s$ corresponds to a "symmetric" or commuting mixture of bosonic and fermionic degrees of freedom. It has been used in [17] for the description of a supersymmetric chain of uncoupled oscillators and it corresponds to the most common choice for combining bosonic and fermionic degrees of freedom. One can find a host of applications, in either problems of physics or mathematics. For instance: in [18–20] we have constructions of coherent states in models described by this algebra; in [16,21] it is applied in the Jaynes-Cummings model; and in [22] in a variant of this model. In [23–30] this algebra is used for studying problems of the representation theory of Lie algebras, Lie superalgebras and their deformations. Some authors [23,31] use the terminology symmetric Clifford-Weyl algebra or Weyl superalgebra. The algebra $W_{as}$ corresponds to an "antisymmetric" or anticommuting mixture of bosons and fermions. Applications—mainly in mathematical problems—can be found in [28,31,32]. Some authors [23,31] refer to this algebra as the antisymmetric Clifford-Weyl algebra.

The Relative Parabose Set $P_{BF}$, the Relative Parafermi Set $P_{FB}$, the Straight Commutation relations $S_{BF}$ and the Straight Anticommutation relations $S_{FB}$ have all been introduced in [5] and constitute different choices of mixing algebraically interacting parabosonic and parafermionic degrees of freedom. Mathematical properties of some of these algebras such as their $G$-graded, $\vartheta$-colored Lie structures and, more generally, their braided group structures have been studied in [12,13] for $P_{FB}$ and in [14,15,33–35] for $P_{BF}$. However, the representation theory of these mixed paraparticle algebras remains an almost unexplored subject. To the best of the author's knowledge, the only works in the bibliography dealing with explicit construction of representations for such algebras has to do with the representations of $P_{BF}^{(1,1)}$ *i.e.*, of the Relative Parabose Set algebra combining a single parabosonic and a single parafermionic degree of freedom [36–38].

Finally, before closing this paragraph and for the sake of completeness, we feel it is worth citing various works appearing in the literature and dealing with algebras which mix particle and paraparticle degrees of freedom (*i.e.*, mixing commutation–anticommutation relations from the above table): One can see for example [39–44] where mainly supersymmetric properties and coherent states are studied for such algebras.

## 3. Braided Group, Ordinary Hopf and $(G, \vartheta)$-Lie Structures for the Mixed Paraparticle Algebras: An Attempt at Classification

### 3.1. Historical and Conceptual Introduction—Literature Review

The notion of $G$-graded Hopf algebra, is not new, either in physics or in mathematics. The idea already appears in some of the early works on Hopf algebras, such as for example in the work of Milnor and Moore [45] where we actually have $\mathbb{Z}$-graded Hopf algebras (see also [46]). It is noteworthy, that such examples initially misled mathematicians to the incorporation of the notion of grading in the definition of the Hopf algebra itself, until about the mid 1960s when P. Cartier and J. Dieudonné removed such restrictions and stated the definition of Hopf algebra in almost its present day form.

Before continuing, we feel it is worth quoting the following proposition which summarizes different conceptual understandings of the notion of the grading of a (complex) algebra $A$ by a finite, Abelian group G (for more details on the following proposition and on the terminology and the notions used in the rest of this section, the interested reader may look at [47–53] and also at Sections 3.3, 3.4, 4.2 of [54]).

**Proposition 3.1:** The following statements are equivalent to each other:

1. $A$ is a $G$-graded algebra (the term superalgebra appears often in physics literature when $G = Z_2$) in the sense that $A = \oplus_{g \in G} A_g$ and $A_g A_h \subseteq A_{gh}$ for any $g, h \in G$.
2. $A$ is a (left) $\mathbb{C}G$-module algebra.
3. $A$ is a (right) $\mathbb{C}G$-comodule algebra.
4. $A$ is an algebra in the Category $_{\mathbb{C}G}\mathfrak{M}$ of representations (modules) of the group Hopf algebra $\mathbb{C}G$.
5. $A$ is an algebra in the Category $\mathfrak{M}^{\mathbb{C}G}$ of corepresentations (comodules) of the group Hopf algebra $\mathbb{C}G$.

We recall here that $A$ being a $\mathbb{C}G$-module algebra is equivalent to saying that $A$ apart from being an algebra is also a $\mathbb{C}G$-module while the structure maps of the algebra (*i.e.*, the multiplication and the unity map which embeds the field into the center of the algebra) are $\mathbb{C}G$-module morphisms (or equivalently homogeneous linear maps whose degree is the neutral element of the group $G$). In the general case of an arbitrary group $G$ the comodule picture would describe the situation more conveniently, however in the above we explicitly use the Hopf algebra isomorphism $\mathbb{C}G \cong (\mathbb{C}G)^*$ between $\mathbb{C}G$ and its dual Hopf algebra $(\mathbb{C}G)^*$ (where $(\mathbb{C}G)^* = Hom(\mathbb{C}G, \mathbb{C}) \cong Map(G, \mathbb{C}) = \mathbb{C}^G$ as complex vector spaces and with $\mathbb{C}^G$ we denote the complex vector space of the set-theoretic maps from the finite abelian group $G$ to $\mathbb{C}$). The essence of the description provided by Proposition 3.1 is that the $G$-grading on the algebra $A$ can be equivalently described as a specific (co)action of the group $G$ (and thus of the group Hopf algebra $\mathbb{C}G$) on $A$ *i.e.*, a (co)action which "preserves" the algebra structure of $A$. Such ideas, which provide an equivalent description of the grading of an algebra $A$ by a group $G$ as a suitable (co)action of the group Hopf algebra $\mathbb{C}G$ on $A$, are actually not new and already appear in works such as [55,56].

What is actually new in the sense that it has been developed since the 1990s and thereafter, is on the one hand the "dualization" of Proposition 3.1 which provides us with the definition of the notion of a "graded coalgebra" and, on the other hand, the role of the notion of the quasitriangularity of the group Hopf algebra $\mathbb{C}G$, in constructing "graded" generalizations of the notion of Hopf algebra itself.

We first collect in the following proposition various alternative readings of the notion of a graded coalgebra:

**Proposition 3.2:** The following statements are equivalent to each other:

1. $C$ is a $G$-graded coalgebra (the term supercoalgebra seems also appropriate when $G = Z_2$) in the sense that $\Delta(C_\kappa) \subseteq \oplus_{g \in G} C_g \otimes C_{g^{-1}\kappa} \equiv \oplus_{gh=\kappa} C_g \otimes C_h$ for any $g, h, \kappa \in G$ and $\varepsilon(C_\kappa) = \{0\}$ for all $\kappa \neq 1 \in G$. ($\Delta : C \to C \otimes C$ and $\varepsilon : C \to \mathbb{C}$ are assumed to be the comultiplication and the counity respectively).
2. $C$ is a (left) $\mathbb{C}G$-module coalgebra.
3. $C$ is a (right) $\mathbb{C}G$-comodule coalgebra.
4. $C$ is a coalgebra in the Category $_{\mathbb{C}G}\mathfrak{M}$ of representations (modules) of the group Hopf algebra $\mathbb{C}G$.

5. $C$ is a coalgebra in the Category $\mathfrak{M}^{\mathbb{C}G}$ of corepresentations (comodules) of the group Hopf algebra $\mathbb{C}G$.

Notice that, in the above proposition, $G$ is considered to be finite and abelian. (See also the proof of the above proposition in the Appendix, for some clarifying comments on the role of these restrictions).

For bibliographic reasons, we should mention at this point, that the notion of a graded coalgebra first appears in the literature in the articles [45,46] and the books [52,53]. However, these references consider the special case for which the grading group is $G = \mathbb{Z}$ and the components of negative degree are zero. To the best of the author's knowledge, the introduction of the notion of graded coalgebra in its full generality, *i.e.*, for an arbitrary grading group $G$, first appears in [57] (where strongly graded coalgebras are also introduced) and is consequently studied in [58–60].

Let us now proceed in briefly describing the way in which the notion of quasitriangularity, its connection with previously known ideas from group theory (e.g., the notion of bicharacter), from Category theory (*i.e.*, the notion of braiding) and its role in the formation of representations and tensor products of graded objects, leads us to direct generalizations of the notion of Hopf algebras and to a novel understanding of the notion of graded Hopf algebras. For what follows, the interested reader on the terminology and the notions of bicharacters, color functions, commutation factors should consult [47,48] and [61–67].

The Universal Enveloping algebras (UEA) of Lie superalgebras (LS) are widely used in physics and they are examples of $\mathbb{Z}_2$-graded Hopf algebras or super-Hopf algebras. These structures strongly resemble Hopf algebras but they are not Hopf algebras themselves, at least not in the ordinary sense. The picture expands even more, if we consider further generalizations of Lie algebras: these are the $\vartheta$-colored $G$-graded Lie algebras or $(G,\vartheta)$-Lie algebras, whose UEAs are $G$-graded Hopf algebras or to be more rigorous $(G,\vartheta)$-Hopf algebras or $G$-graded, $\vartheta$-braided Hopf algebras (see the relative discussion in [35,47]). In this last case, $\vartheta : G \times G \to \mathbb{C}^*$ stands for a skew-symmetric bicharacter [47] on $G$ (or: commutation factor [61–63] or color function [65,66]), which has been shown [47,64] to be equivalent to a triangular universal $R$-matrix on the group Hopf algebra $\mathbb{C}G$. This finally entails [47–49,64] a symmetric braiding in the Monoidal Category $_{\mathbb{C}G}\mathfrak{M}$ of the modules over the group Hopf algebra $\mathbb{C}G$.

In fact, in [47,64] a simple bijection is described, from the set of bicharacters of a finite abelian group $G$ onto the set of Universal $R$-matrices of the group Hopf algebra $\mathbb{C}G$ [64] and from there onto the set of the braidings of the monoidal Category of representations $_{\mathbb{C}G}\mathfrak{M}$ ([47], Theorem 10.4.2)

In other words  Bicharacters

$$
\begin{array}{ccc}
\text{Bicharacters} & \xleftarrow{\ \text{"1-1"}\ }\xrightarrow{\ } \text{Universal } R\text{-matrices} & \xleftarrow{\ \text{"1-1"}\ }\xrightarrow{\ } \text{Braidings for the} \\
\text{on G} & \text{on } \mathbb{C}G & _{\mathbb{C}G}\mathfrak{M}\ \text{Category}
\end{array}
$$

The correspondence is such that given a bicharacter $\vartheta : G \times G \to \mathbb{C}^*$, the corresponding $R$-matrix is given by [64]

$$
R = \sum R^{(1)} \otimes R^{(2)} = \frac{1}{n^2} \sum_{\substack{g,h \in G \\ g',h' \in G'}} \vartheta(g,h)\overline{\langle g',g\rangle}\overline{\langle h',h\rangle}\, g' \otimes h'
$$

and the corresponding braiding of the monoidal Category of representations ${}_{\mathbb{C}G}\mathfrak{M}$, by the family of isomorphisms $\psi_{V,W}: V \otimes W \to W \otimes V$ given by $\psi_{V,W}(x \otimes y) = \sum R^{(2)} \cdot y \otimes R^{(1)} \cdot x = \vartheta(g,h)\, y \otimes x$ for any $x \in V_g, y \in W_h; g, h \in G$. In the above, we have denoted by $\overline{c}$ the complex conjugate of any complex number $c$, by $\mathbb{C}^*$ the multiplicative group of non-zero complex numbers, by $G'$ the character group of $G$ and by $\langle\ ,\ \rangle: G' \times G \to \mathbb{C}^*$ the canonical pairing $\langle g', g \rangle = g'(g) \in \mathbb{C}^*$ for all $g' \in G', g \in G$. The vector spaces $V, W$ are any two $\mathbb{C}G$-modules *i.e.* any two $G$-graded vector spaces and by "·" we denote the action of the group elements on the elements of the corresponding vector space. The above described bijection is such that [64] the skew-symmetric bicharacters (*i.e.*, the color functions or commutation factors) are mapped onto triangular universal $R$-matrices and thus onto symmetric braidings of ${}_{\mathbb{C}G}\mathfrak{M}$ (see also Sections 3.5.3 and 4.2 of [54] for detailed calculations for the simplest example of $\mathbb{C}Z_2$). Also, recall that a character $\chi$ of $G$ is a homomorphism $\chi: G \to \mathbb{C}^*$ of $G$ to the multiplicative group of non-zero complex numbers $\mathbb{C}^* = (\mathbb{C} \backslash \{0\}, \times)$, *i.e.*, $\chi(gh) = \chi(g)\chi(h), \forall g, h \in G$ and that the characters form a (multiplicative) group $G'$, which in the finite, abelian case is isomorphic to $G$ *i.e.*, $G \cong G'$ as abelian groups and thus $\mathbb{C}(G') \cong \mathbb{C}G \cong (\mathbb{C}G)^*$ as Hopf algebras.

According to the modern terminology [47–49,64] developed in the 1990s and originating from the Quantum Groups theory, $(G, \vartheta)$-Hopf algebras belong to the—conceptually wider—class of Braided Groups (in the sense of the braiding described above). Here we use the term "braided group" loosely, in the sense of [48,49]. It is also customary to speak of such structures as Hopf algebras in the braided Monoidal Categories ${}_{\mathbb{C}G}\mathfrak{M}$ of representations of $\mathbb{C}G$. The following proposition (see [47–49]) summarizes various different conceptual understandings of the term $G$-graded, $\vartheta$-braided Hopf algebra (see also the corresponding definitions of [47–49,68]).

**Proposition 3.2:** The following statements are equivalent to each other:

1. $H$ is a $G$-graded, $\vartheta$-braided Hopf algebra or a $(G, \vartheta)$-Hopf algebra.

2. $H$ is a Hopf algebra in the braided Monoidal Category ${}_{\mathbb{C}G}\mathfrak{M}$ of representations of $\mathbb{C}G$.

3. $H$ is a braided group for which the braiding is given by the function $\vartheta: G \times G \to \mathbb{C}^*$.

4. $H$ is simultaneously an algebra, a coalgebra and a $\mathbb{C}G$-module, all its structure functions (multiplication, comultiplication, unity, counity and antipode) are $\mathbb{C}G$-module morphisms. The comultiplication $\underline{\Delta}: H \to H \underline{\otimes} H$ and the counity $\underline{\varepsilon}: H \to \mathbb{C}$ are algebra morphisms in the braided monoidal Category ${}_{\mathbb{C}G}\mathfrak{M}$. ($H \underline{\otimes} H$ stands for the braided tensor product algebra). At the same time, the antipode $S: H \to H$ is a "twisted" or "braided" anti-homomorphism in the sense that $S(xy) = \vartheta(\deg(x), \deg(y))S(y)S(x)$ for any homogeneous $x, y \in H$.

5. The $\mathbb{C}G$-module $H$ is an algebra in ${}_{\mathbb{C}G}\mathfrak{M}$ (equiv.: a $\mathbb{C}G$-module algebra) and a coalgebra in ${}_{\mathbb{C}G}\mathfrak{M}$ (equiv.: a $\mathbb{C}G$-module coalgebra), the comultiplication $\underline{\Delta}: H \to H \underline{\otimes} H$ and the counity $\underline{\varepsilon}: H \to \mathbb{C}$ are algebra morphisms in the braided monoidal Category ${}_{\mathbb{C}G}\mathfrak{M}$ and at the same time, the antipode $S: H \to H$ is an algebra anti-homomorphism in the braided monoidal Category ${}_{\mathbb{C}G}\mathfrak{M}$.

The investigation of such structures for the case of the paraparticle algebras has been an old issue: The "free" Parafermionic $P_F$ and parabosonic $P_B$ algebras have been shown to be (see the discussion in the introduction, in Section 2 and also [69,70] for a review) isomorphic to the Universal Enveloping Algebra (UEA) of a Lie algebra and a Lie superalgebra (or: $\mathbb{Z}_2$-graded Lie algebra) respectively, while the Relative Parabose set algebra $P_{BF}$ has been shown [14,15] to be isomorphic to the UEA of a $(\mathbb{Z}_2 \times \mathbb{Z}_2)$-graded Lie algebra. At the same time the Relative Parafermi set algebra $P_{FB}$ has been shown [12,13] to be isomorphic to the UEA of a Lie superalgebra. In [69,71] we have studied the case of $P_B$, and we establish its braided group structure (here: $\mathbb{Z}_2$-graded Hopf structure) independently of its $\mathbb{Z}_2$-graded Lie structure.

### 3.2. Description of the Problem–Research Objectives

At this point, we feel it will be quite useful to try to shed some light on the following subtle points, which lie at the heart of our proposed investigation:

On the one hand, speaking about a single $G$-graded algebra $A$, there may—in principle—exist more than a single braided group structure that can be attached to it. In other words, given a specific $G$-grading, the (corresponding) braiding is not necessarily unique. This can be seen in some simple examples, maybe even for some cases of UEAs of $\vartheta$-colored $G$-graded Lie algebras: since the symmetric braidings are in a bijective correspondence [47,64] with the skew-symmetric bicharacters (on the finite abelian group $G$) or with the triangular universal R-matrices (of the corresponding group Hopf algebra $\mathbb{C}G$), we can easily see that even for the case of a single $(\mathbb{Z}_2 \times \mathbb{Z}_2)$-graded associative algebra, there may—in principle—exist different ($\mathbb{Z}_2 \times \mathbb{Z}_2$)-graded Hopf algebras (*i.e.*, braided groups) corresponding to it. The difference stems from the possibility to pick different braidings (*i.e.*, different colors or different commutation factors) for the finite, abelian $\mathbb{Z}_2 \times \mathbb{Z}_2$ group (see also [67] for examples on the available possibilities of such choices) and reflects on the differentiation in the definitions of the comultiplication $\Delta : A \to A \underline{\otimes} A$ and the antipode $S : A \to A^{gr.op}$ ($A^{gr.op}$ is the graded-opposite algebra). Conceptually (in the language of Category Theory), we may equivalently say that, the difference stems from the possibility to pick different (non-trivial) R-matrices for the $\mathbb{C}(\mathbb{Z}_2 \times \mathbb{Z}_2)$ group Hopf algebra and reflects on different families of permuting isomorphisms (braidings) between the tensor product representations of the $(\mathbb{Z}_2 \times \mathbb{Z}_2)$-graded $A$-modules and between the tensor powers of $A$ itself.

On the other hand, the picture may become even more complicated by the fact that the $G$-grading for $A$, is not uniquely assigned itself: In other words, for a single algebra $A$, there may exist group-gradings by different groups and even if we consider a single group G, it may assign non-equivalent gradings to the same algebra $A$. In order to elucidate this last point we recall here, that it has been shown [56] that a concrete $G$-grading on the $\Bbbk$-algebra $A$, is equivalent to a concrete $\Bbbk G$-(co)action on $A$. Consequently, the problem of classifying all the possible gradings induced by G on $A$ is equivalent to classifying all the (non-isomorphic) $\Bbbk G$-(co)module algebras which are all the (non-isomorphic) $\Bbbk G$-(co)modules with carrier space $A$, whose (co)action preserves the algebra structure of $A$ (in the sense of Proposition 3.1).

In [34,35,37,38] we have already started a preliminary investigation of some of the above points, for the case of the Relative Parabose Set algebra $P_{BF}$: In [34,35] we review $P_{BF}$ as the UEA of a $(\mathbb{Z}_2 \times \mathbb{Z}_2)$-graded, $\vartheta$-colored Lie algebra (for a specific choice of the commutation factor $\vartheta$ proposed in [14,15]). However, in [37,38] we adopt a different point of view, in which we consider $P_{BF}$ as a $(\mathbb{Z}_2 \times \mathbb{Z}_2)$-graded associative algebra, with a different (inequivalent) form of the grading $i.e.$, with a different $\mathbb{C}(\mathbb{Z}_2 \times \mathbb{Z}_2)$-action. In this last case, the $(\mathbb{Z}_2 \times \mathbb{Z}_2)$-grading is not necessarily associated to some particular color-graded Lie structure. We intend to rigorously investigate further, the following points:

- Given the $(\mathbb{Z}_2 \times \mathbb{Z}_2)$-grading described in [14,15,34,35] we intend to check whether it is compatible with other commutation factors $\vartheta$ ($i.e.$,: other braidings for the $_{\mathbb{C}(\mathbb{Z}_2 \times \mathbb{Z}_2)}\mathfrak{M}$ Category of modules) than the one presented in these works. In other words, we are going to determine possible alternative braided group structures, corresponding to the single $(\mathbb{Z}_2 \times \mathbb{Z}_2)$-graded structure for $P_{BF}$ described in the above works. It will also be interesting to examine, which of these alternatives—if any—are directly associated to some particular color-graded Lie structure (directly in the sense that they may stem from the UEA).

- We are going to determine possible alternative $G$-gradings for the $P_{BF}$, $P_{FB}$ (co)algebras where the group $G$ may either be $\mathbb{Z}_2 \times \mathbb{Z}_2$ itself (with some grading inequivalent to the previous, in the sense formerly described) or some other suitable group, for ex. $\mathbb{Z}_2$ or $\mathbb{Z}_4$. In each case, we will further investigate the possible braidings (in the sense analyzed in the former paragraph).

- We are going to collect the results of the previous two steps and develop Theorems and Propositions which establish the possible braided group structures of $P_{BF}$ and $P_{FB}$ independently of the possible color-graded Lie structures. For each of the above cases, we intend to explicitly compute: (a) The group action ($i.e.$, the grading); (b) The braiding ($i.e.$, the family of isomorphisms), the commutation factor ($i.e.$, the bicharacter or equiv: the color function), (c) The (quasi)triangular structure ($i.e.$, the $R$-matrix) of the corresponding group Hopf algebra.

- Finally, in each of the above cases we intend to apply bosonization [48,72] or bosonization-like techniques (in the sense we have done so in [69–71]) to obtain ordinary Hopf structures (with no grading and with trivial braiding) with equivalent representation theories.

We can finally summarize the above discussion in three research objectives:

**1st Research Objective:** The first problem we intend to investigate is the classification of the gradings induced on the paraparticle (co)algebras (especially on $P_{BF}$ and $P_{FB}$ algebras) by small order finite Abelian Groups such as $Z_2, Z_3, Z_4, Z_2 \times Z_2$ $etc$. In other words, we intend to classify those group (co)actions which preserve the corresponding (co)algebra structures, turning thus the (co)algebras in $\mathbb{C}G$-(co)module (co)algebras.

Let us also mention at this point, that similar problems of investigating and classifying the gradings induced on various different algebras by a group $G$, have received much attention during the last decade. Far from trying to present an exhaustive bibliography at this point we feel it is worth mentioning some references indicating the breadth of the associated problems: In [73–76] gradings on various matrix algebras are investigated, in [77–95] we have results on studies, properties and classifications for gradings on different kinds of Lie algebras and in [96–99] gradings on various different associative and non-associative algebras are examined.

**2nd Research Objective:** Further, for each of the above gradings we intend to classify the corresponding braided group structures. In other words, we will write down the possible bicharacters of the above groups or equivalently the possible $R$-matrices of the corresponding group Hopf algebras or equivalently the braidings of the corresponding Category $_{CG}\mathfrak{M}$ (or $\mathfrak{M}^{CG}$) of modules (or comodules). For each one of these braidings, we aim to examine whether or not there are available compatible graded algebraic and coalgebraic structures suitable for producing a braided group.

Studies dealing with classifications of R-matrices and braidings and which seem to be related to the proposed idea can be found in [100,101] (see also [102]).

**3rd Research Objective:** Apply or develop suitable bosonization or bosonization-like techniques to obtain ordinary Hopf structures, with no grading and with trivial braiding, possessing equivalent representation theories.

## 4. An Attempt to Approach the Fock-like Representations for the $P_B, P_F, P_{BF}, P_{FB}$ Algebras Utilizing Their Braided Group Structures

### 4.1. Conceptual Introduction–Methodological Review

In [103], we take advantage of the super-Hopf structure of $P_B$ which has been extensively studied in [69–71], and based on it, we develop a "braided interpretation" of the Green ansatz for parabosons. We further develop a method, for employing this braided interpretation in order to construct analytic expressions for the matrix elements of the Fock-like representations of $P_B$. Concisely, the method consists of the following steps:

➢ regarding $CAR$ (the usual Weyl algebra or: boson algebra) as a superalgebra with odd generators, and proving that it is isomorphic (as an assoc. superalgebra) to a quotient superalgebra of $P_B$,

➢ constructing the graded tensor product representations, of (graded) tensor powers of the form $CAR \otimes CAR \otimes ... \otimes CAR$ ($p$-copies),

➢ pulling back the module structure to a representation of $P_B$ through suitable (homogeneous) homomorphisms of the form $P_B \rightarrow CAR \otimes CAR \otimes ... \otimes CAR$, which are constructed via the braided comultiplication $\Delta : P_B \rightarrow P_B \underline{\otimes} P_B$ of $P_B$ (see [103]),

➢ prove that the $P_B$-modules thus obtained, are isomorphic (as $P_B$-modules) to $\mathbb{Z}_2$-graded tensor product modules, between $p$-copies, of the first ($p=1$) Fock-like representation of $P_B$,

➢ prove that the parabosonic $p$-Fock-like module, corresponding to arbitrary value of the positive integer $p$, is contained as an irreducible direct summand of the above constructed $\mathbb{Z}_2$-graded tensor product representation,

➢ compute explicitly the action of the $P_B$ generators and the corresponding matrix elements, on the above mentioned $p$-Fock-like modules and finally,

➢ decompose the obtained $\mathbb{Z}_2$-graded tensor product representations into irreducible components and investigate whether more irreducible summands arise, non-isomorphic to the $p$-Fock-like submodule.

*4.2. Description of the Problem–Research Objectives*

The—possible—advantage of the formerly described method, is that it may permit us to explicitly construct unitary, irreducible representations (unirreps) with general lowest weight vectors of the form $(p_1, p_2,...)$. However, we must mention at this point that the application of the above method in [103] has not been finalized due to computational difficulties encountered and which will be described in the sequel. Consequently, the research objectives of this part of the project consist of refining, applying and generalizing the above method:

• We first intend to proceed to the explicit construction of the Fock-like representations in the case of the (inf. deg. of freedom) parabosonic $P_B$ and parafermionic $P_F$ algebra following the methodology developed in [103] and outlined above. Starting from the parabosonic algebra, this involves computations of expressions of the following form

$$\prod_{r,i_r=1}^{\infty} (B_{i_r}^+)^{n_{r,i_r}} \rhd |0\rangle = \prod_{r,i_r=1}^{\infty} (\sum_{k=1}^{p} b_{i_r}^{(k)+})^{n_{r,i_r}} \rhd |0\rangle = \prod_{r,i_r=1}^{\infty} (\sum_{k=1}^{p} I \otimes I \otimes ... \otimes b_{i_r}^+ \otimes I \otimes ... \otimes I)^{n_{r,i_r}} \rhd |0\rangle$$

where: $b_{i_r}^{(k)+} = I \otimes I \otimes ... \otimes b_{i_r}^+ \otimes I \otimes ... \otimes I$, $\rhd$ denotes the action, $|0\rangle \equiv |0\rangle \otimes |0\rangle \otimes ... \otimes |0\rangle$ the $p$-fold tensor product of the bosonic ground state, the CCR generator $b_{i_r}^+$ lies in the $k$-th entry of the tensor product and there are a finite only number of non-zero exponents $n_{r,i_r}$ in the above product. The mathematical problem here, which is necessary to be solved in order to explicitly perform the computation is the development of a suitable multinomial theorem in the anticommuting variables $b_{i_r}^{(k)+}$. The corresponding problem appears to be easier for the case of $P_F$, since the corresponding variables $f_{i_r}^{(k)+} = I \otimes I \otimes ... \otimes f_{i_r}^+ \otimes I \otimes ... \otimes I$ ($f_{i_r}^+$ is the CAR generator) appear to be commuting (the exact choice of the braiding and the grading depends of course on the results of the previous part of the project). What we are actually describing here, are the steps for the explicit calculation of the action of the generators on the tensor product representations of—suitably—graded versions of CCR and CAR and the subsequent decomposition of these representations in irreducible components. In [103] we have proved that the $p$-Fock-like modules are contained as irreducible factors of such graded, tensor product representations. However, it remains to see whether such decompositions can produce as direct summands or more generally as submodules other non-equivalent representations as well.

Before proceeding with the discussion, we summarize in the following table the present state of knowledge about the parabosonic Fock-like representations, including the previous discussion (In the following table $b_i^+$ and $B_j^+$ denote the CCR and the $P_B$ generators respectively and m denotes the number of the generators i.e. the possible values of $i$ and $j$):

## Boson (CCR) and paraboson ($P_B$) representations:

| | |
|---|---|
| **$m = 1, p = 1$**<br><br>single particle<br><br>Bosonic (CCR) Fock representation<br><br>$$\lvert n \rangle = \frac{(b^+)^n}{\sqrt{n!}} \lvert 0 \rangle$$<br><br>• This is the celebrated Heisenberg-Schröedinger representation, leading to the matrix mechanics or the wave mechanics formulation of elementary QM<br><br>• The wave mechanical description is provided by the Hermite polynomials (times a suitable exponential decay factor) | **$m > 1, p = 1$**<br><br>multi particle<br><br>Bosonic (CCR) Fock representation<br><br>$$\lvert n_1, \dots, n_i, \dots \rangle = \frac{(b_1^+)^{n_1} \cdots (b_i^+)^{n_i} \cdots}{\sqrt{n_1! \cdots n_i! \cdots}} \lvert 0 \rangle$$<br><br>• This is known as the Fock or the Fock-Cook representation<br><br>• It can be constructed by forming the ordinary (ungraded) tensor product of the $n = 1$ case (see the previous column)<br><br>• This is the mathematical basis on which the QFT elaborates |
| **$m = 1, p > 1$**<br><br>single particle<br><br>Parabonic ($P_B$) Fock-like representation<br><br>$$\lvert 2n \rangle = \frac{(B^+)^{2n}}{2^n \sqrt{n! \left( {\tfrac{p}{2}} \right)_n}} \lvert 0 \rangle$$<br><br>$$\lvert 2n+1 \rangle = \frac{(B^+)^{2n+1}}{2^n \sqrt{n! 2 \left( {\tfrac{p}{2}} \right)_{n+1}}} \lvert 0 \rangle$$<br><br>• For the wave mechanical description see Yang [104] (1951) and Ohnuki [105], Sharma [106] (1978) | **$m > 1, p > 1$**<br><br>multi particle<br><br>Parabosonic ($P_B$) Fock-like representation<br><br>$$P(B_i^+) \cdot \lvert 0 \rangle = \prod_{r, i_r = 1}^{\infty} (B_{i_r}^+)^{n_{r, i_r}} \cdot \lvert 0 \rangle =$$<br><br>$$\prod_{r, i_r = 1}^{\infty} \left( \sum_{k=1}^{p} b_{i_r}^{(k)+} \right)^{n_{r, i_r}} \cdot \lvert 0 \rangle =$$<br><br>$$= \prod_{r, i_r = 1}^{\infty} \left( \sum_{k=1}^{p} I \otimes \dots \otimes b_{i_r}^+ \otimes \dots \otimes I \right)^{n_{r, i_r}} \cdot \lvert 0 \rangle = ?$$<br><br>• Explicit construction (matrix elements, formulae for the action of the generators etc) for the general case of the infinite degrees of freedom has been given by Lievens, Stoilova, van der Jeugt [6–8] (2007–2008) |

• Next, we intend to compare our obtained (according to the above described method) results with those obtained in [6–8] (where a totally different approach, based on induced representations and chains of inclusions of Lie superalgebras contained as subalgebras, has been adopted). It is expected that the identification of the representations may lead us to valuable insight, relative to the interrelations between the various, diversified analytical tools used.

• The next step will consist of generalizing the above calculations for the case of the mixed paraparticle algebras $P_{BF}$ and $P_{FB}$. The philosophy of the method is based on the same idea: The Fock-like representations of $P_{BF}$ and $P_{FB}$ will be extracted as irreducible submodules arising in the decomposition of the graded tensor product representations of $W_s$ and $W_{as}$. In this case, $W_s$ is a mixture of commuting (symmetric mixture) bosons and fermions and $W_{as}$ a mixture of anticommuting (antisymmetric mixture) of bosonic and fermionic generators (see also [54] § 6.2 pp. 199–207, [31] for more details on the structure of these algebras). Just as the CCR may be considered a graded quotient algebra of $P_B$ (see [103]) , and the CAR a graded quotient algebra of $P_F$, in the same spirit we will consider $W_s$ as a suitable graded quotient of $P_{BF}$ and $W_{as}$ as a graded quotient of $P_{FB}$. These are exactly the algebras we intend to employ, in order to generalize the formerly described method for the case of the mixed paraparticle algebras $P_{BF}$ (Relative Parabose Set algebra) and $P_{FB}$ (Relative Parafermi Set algebra). The results of the previous part of the project (*i.e.*, Section 3.) are expected to lead us in suitable choices for the grading and the braiding of $W_s$ and $W_{as}$ (in the same manner that the results of [69–71] led us to the use of odd-bosons in [103]). Finally it is worth mentioning, that the computational problem we expect to reveal here is the development of a suitable multinomial theorem mixing commuting and anticommuting variables.

## 5. A Proposal for the Development of an Algebraic Model for the Description of the Interaction between Monochromatic Radiation and a Multiple Level System

### 5.1. Review of Recent Work

In [34,35] (see also [107]) we have studied algebraic properties of the Relative Parabose algebra $P_{BF}$ and the Relative Parafermi algebra $P_{FB}$ such as their gradings, braided group structures, θ-colored Lie structures, their subalgebras, *etc*. These algebras, constitute paraparticle systems defined in terms of parabosonic and parafermionic generators (or: interacting parabosonic and parafermionic degrees of freedom, in a language more suitable for physicists) and trilinear relations. We have then proceeded in building realizations of an arbitrary Lie superalgebra $L = L_0 \oplus L_1$ (of either fin or infin dimension) in terms of these mixed paraparticle algebras. Utilizing a given $\mathbb{Z}_2$-graded, finite dimensional, matrix representation of $L$, we have actually constructed maps of the form $J : L \to gl(m/n) \subset \genfrac{}{}{0pt}{}{P_{BF}}{P_{FB}}$ from the LS $L$ into a copy of the general linear superalgebra $gl(m/n)$ isomorphically embedded into either $P_{BF}$ or into $P_{FB}$. These maps have been shown to be graded Hopf algebra homomorphisms or more generally braided group homomorphisms and constitute generalizations and extensions of older results [107]. From the viewpoint of mathematical physics, these maps generalize—in various aspects (see the discussion in [35])—the standard bosonic-fermionic Jordan-Scwinger [108,109] realizations of Quantum mechanics. In [37,38] we have further proceeded in building and studying a class of irreducible representations for the simplest case of the $P_{BF}^{(1,1)}$ algebra in a single parabosonic and a single parafermionic degree of freedom (a 4-generator algebra). We have used the terminology "Fock-like representations" because these representations apparently generalize the well known boson-fermion Fock spaces of Quantum Field theory.

The carrier spaces of the Fock-like representations of $P_{BF}^{(1,1)}$ constitute a family parameterized by the values of a positive integer $p$. They have the general form $\bigoplus_{n=0}^{p} \bigoplus_{m=0}^{\infty} V_{m,n}$ where $p$ is an arbitrary (but fixed) positive integer. The subspaces $V_{m,n}$ are 2-dim except for the cases $m = 0$, $n = 0$, $p$, *i.e.*, except the subspaces $V_{0,n}$, $V_{m,0}$, $V_{m,p}$ which are 1-dim for all values of $m$ and $n$. These subspaces can be visualized as follows:

| $V_{0,0}$ | $V_{0,1}$ | ... | $V_{0,n}$ | ... | ... | $V_{0,p-1}$ | $V_{0,p}$ |
| $V_{1,0}$ | $\boldsymbol{V_{1,1}}$ | ... | $\boldsymbol{V_{1,n}}$ | ... | ... | $\boldsymbol{V_{1,p-1}}$ | $V_{1,p}$ |
| $\vdots$ | $\vdots$ | $\ddots$ | $\vdots$ | $\vdots$ | $\ddots$ | $\vdots$ | $\vdots$ |
| $V_{m,0}$ | $\boldsymbol{V_{m,1}}$ | ... | $\boldsymbol{V_{m,n}}$ | $\boldsymbol{V_{m,n+1}}$ | ... | $\boldsymbol{V_{m,p-1}}$ | $V_{m,p}$ |
| $\vdots$ | $\vdots$ | ... | $\boldsymbol{V_{m+1,n}}$ | ... | ... | $\vdots$ | $\vdots$ |
| $\vdots$ | $\vdots$ | $\ddots$ | $\vdots$ | ... | $\ddots$ | $\vdots$ | $\vdots$ |

Notice that in the above figure, the subspaces of the first and the $p$-th column as well as the subspaces of the first row correspond to 1d subspaces while the "inner" subspaces (which are bold in the figure) correspond to 2d subspaces. The generators $b^+, b^-, f^+, f^-$ of $P_{BF}^{(1,1)}$ are acting (see [37] for details) as creation-annihilation operators on the above "two"-dimensional ladder of subspaces: The action of the $b^+ (b^-)$ operators produces upward (downward) vertical shifts, changing thus the value of the line, while the action of the $f^+ (f^-)$ operators produces right (left) shifts, changing thus the value of the columns. Finally, note that the action of the $f^+$ operator, on the above described vector space, is a nilpotent one satisfying $(f^+)^{p+1} = 0$ (for the corresponding representation characterized by this specific value of $p$).

### 5.2. Description of the Problem–Research Objectives

Our research objective has to do with a potential physical application of the of the paraparticle and LS Fock-like representations discussed above, in the extension of the study of a well-known model of quantum optics: The Jaynes-Cummings model [16] is a fully quantized—and yet analytically solvable—model describing (in its initial form) the interaction of a monochromatic electromagnetic field with a two-level atom. Using the Fock-like modules described above, we will attempt to proceed in a generalization of the above model in the study of the interaction of a monochromatic parabosonic field with a $(p + 1)$-level system. The Hamiltonian for such a system might be of the form

$$H_{dyn} = H_b + H_f + H_{interact} = \frac{\omega_b}{2}\left\{b^+, b^-\right\} + \frac{\omega_f}{2}\left[f^+, f^-\right] + \frac{(\omega_f - \omega_b)p}{2} + \frac{\lambda}{2}\left(\left\{b^-, f^+\right\} + \left\{b^+, f^-\right\}\right)$$

Or more generally:

$$H_{dyn}^{\,*} = H_b + H_f + H_{interact}^{*} = \frac{\omega_b}{2}\left\{b^+, b^-\right\} + \frac{\omega_f}{2}\left[f^+, f^-\right] + \frac{(\omega_f - \omega_b)p}{2} + \lambda_1 b^- f^+ + \lambda_2 f^+ b^- + \lambda_2^* b^+ f^- + \lambda_1^* f^- b^+$$

where $\omega_b$ stands for the energy of any paraboson field quanta (this generalizes the photon, represented by the Weyl algebra part of the usual JC-model), $\omega_f$ for the energy gap between the subspaces $V_{m,n}$ and $V_{m,n+1}$ (this generalizes the two-level atom, represented by the su(2) generators of the usual JC-model) and $\lambda$ or $\lambda_i$ ($i = 1,2$) suitably chosen coupling constants. Notice that $\omega_b$ and $\omega_f$ might be some functions of m or n or both. The $H_b + H_f$ part of the above Hamiltonians represents the "field" and the "atom" respectively, while the $H_{interact} = \frac{\lambda}{2}\left(\left\{b^-, f^+\right\} + \left\{b^+, f^-\right\}\right)$, $H_{interact}^{*} = \lambda_1 b^- f^+ + \lambda_2 f^+ b^- + \lambda_2^* b^+ f^- + \lambda_1^* f^- b^+$ operators "simulate" the "field-atom" interactions causing transitions from any $V_{m,n}$ subspace to the subspace $V_{m-1,n+1} \oplus V_{m+1,n-1}$ (absorptions and emissions of radiation). The Fock-like representations, the formulas for the action of the generators and the corresponding carrier spaces, will provide a full arsenal for performing actual computations in the above conjectured Hamiltonian and for deriving expected and mean values for desired physical quantities. A preliminary version of these ideas, for the simplest case of $P_{BF}^{(1,1)}$ has already appeared (see the discussion at Section 5 of [37]). The spectrum generating algebra of $H$ may be considered to be either $P_{BF}^{(1,1)}$ or $P_{FB}^{(1,1)}$ or more generally any other mixed paraparticle algebra whose representations can be directly deduced from those of $P_{BF}^{(1,1)}$ or $P_{FB}^{(1,1)}$: Such algebras may be the "straight" Paraparticle algebras $SCR^{(1,1)} \cong P_B^{(1)} \otimes^{Gr} P_F^{(1)}$ or $SAR^{(1,1)} \cong P_B^{(1)} \otimes_{gr} P_F^{(1)}$ where $\otimes^{Gr}$ and $\otimes_{gr}$ stand for braided tensor products for suitable choices of the grading group $G$ and the braiding function $\theta$. More details on the choices of the grading groups and the braiding functions and on the above mentioned isomorphisms will be given in the forthcoming work [110].

In this way, we will actually construct a family of exactly solvable, quantum mechanical models, whose properties will be studied quantitatively (computation of energy levels, eigenfunctions, rates of transitions between states, *etc.*) and directly compared with theoretical and experimental results.

Last, but not least, it is expected that the study of such models will provide us with deep insight into the process of Quantization itself: We will be able to proceed in direct comparison between mainstream quantization methods of Quantum Mechanics where the operators representing the interaction, *i.e.*, the dynamics of the system, are explicitly contained as summands of the form $H_{interact} = \dfrac{\lambda}{2}\left(\{b^-, f^+\} + \{b^+, f^-\}\right)$, or $H_{interact}^* = \lambda_1 b^- f^+ + \lambda_2 f^+ b^- + \lambda_2^* b^+ f^- + \lambda_1^* f^- b^+$ of the Hamiltonian, and the idea of Algebraic (or Statistical) Quantization as this is outlined in works such as [111]: In this case, the idea is to exploit "free" Hamiltonians of the form

$$H_{free} = H_b + H_f = \frac{\omega_b}{2}\left\{b^+, b^-\right\} + \frac{\omega_f}{2}\left[f^+, f^-\right] + \frac{(\omega_f - \omega_b)p}{2}$$

which contain no explicit dynamical interaction terms but include the interaction implicitly into the relations of the spectrum generating algebra itself. Since the spectrum generating algebra can be chosen among $P_{BF}^{(1,1)}, P_{FB}^{(1,1)}, SCR^{(1,1)}, SAR^{(1,1)}$, and its corresponding representation by fixing a concrete value for the positive integer p, we can have a multitude of models of this form which deserve to be further investigated. It is "natural" to start by studying more conventional Hamiltonians of the form $H_{dyn}, H_{dyn}^*$ using as spectrum generating algebras either $SCR^{(1,1)}$ or $SAR^{(1,1)}$ or to use the "free" Hamiltonian $H_{free}$ in combination with a spectrum generating algebra such as $P_{BF}^{(1,1)}$ or $P_{FB}^{(1,1)}$, without of course excluding all the other possibilities as well (using for example $P_{BF}^{(1,1)}$ in conjunction with either $H_{dyn}$ or $H_{dyn}^*$). The reason for this preference can be well understood if one takes a look at the description of these algebras given in the table of Section 2 in terms of generators and relations: the multitude of the algebraic relations of the "relative" set algebras $P_{BF}$ or $P_{FB}$ in contrast to the SCR and SAR algebras where only commutation (anticommutation) relations are involved between generators of different "species" indicate that we may expect a more promising simulation of the dynamics by the $P_{BF}$ or $P_{FB}$ algebras in conjunction with the "free" Hamiltonian $H_{free}$.

We intend to come back shortly with more details and the first results of the above ideas.

## 6. Conclusions

We have reviewed certain aspects of the mathematical theory of the various paraparticle algebras in an attempt to outline three distinct branches of a long-term project aimed at: (a) the study of structural properties such as the classification of the various gradings and braided group structures of these algebras; (b) the explicit construction of classes of representations, utilizing different gradings and braided group structures; and (c) the investigation of the usefulness of these algebras in modeling the interaction of a monochromatic field with a multiple level system.

After the introduction in Section 1, where a brief historical review is made of the most important developments in the mathematical study of these algebras, we proceed in Section 2 to the introduction of the family of algebras we are going to discuss, in terms of generators and relations.

In Section 3, after a conceptual introduction to the modern algebraic treatment of notions such as grading, brainding, bicharacters, color functions, commutation factors and the role of the quasitriangular group Hopf algebras in building this understanding, the investigation is focused on the classification of the various possible actions of low-order abelian groups on the paraparticle algebras and the classification of the various R-matrices for these groups.

In Section 4, a method is proposed, based on the use of braided tensor products of representations of CCR, CAR, $W_s$ and $W_{as}$ for the explicit construction of families of Fock-like representations of the paraparticle algebras. Special attention is paid in the description of unsolved mathematical problems related to the method and dealing with the development of multinomial expansions mixing commuting and anticommuting variables.

Finally, in Section 5, we propose a family of Hamiltonians built on paraparticle degrees of freedom together with families of corresponding Fock-like representations, and discuss their suitability in the description of the radiation–matter interaction via paraparticle generalizations of the celebrated Jaynes-Cummings model of Quantum Optics.

## Acknowledgments

Part of this work was initialized during the second half of 2010 and the beginning of 2011 while the author was a postdoctoral fellow researcher, supported by CONACYT/J60060, at the Institute of Physics and Mathematics (IFM) of the University of Michoacan (UMSNH) at Morelia, Michoacan, Mexico. Since the summer of 2011, the author has been a postdoctoral researcher at the School of Mathematics of the Aristotle University of Thessaloniki (AUTH) at Thessaloniki Greece, supported by a research fellowship granted from the Research Committee of AUTH. The author acknowledges the financial support from both sources and wishes to express his gratitude to the staff of both institutions for their kind and valuable support. He would like to thank especially Alfredo Herrera-Aguilar and Costas Daskaloyannis for organizing seminars at UMSNH and AUTH respectively, where different parts of the above work were presented and for inspiring discussions relative to the content of the present paper.

**Appendix: Sketch of the Proof of Proposition 3.2**

We will not provide here a full proof of Proposition 3.2, as this (together with a detailed description of the terminology involved) would require quite a lot of space and would go outside the scope of this paper. We will however give a detailed proof of the implication: $3. \Rightarrow 1.$ in order to provide a taste of "what's really going on". The interested reader can surf through the references provided in § 3 of the main body of the article.

_Proof of the implication_ $3. \Rightarrow 1.$ _of Proposition 3.2:_

Let us first begin with some preliminary facts: If $H$ is a Hopf algebra and $B, C$ are two (right) $H$-comodules through $\rho_B : B \to B \otimes H$ written explicitly: $\rho_B(b) = \sum b_0 \otimes b_1$ and $\rho_C : C \to C \otimes H$ written explicitly: $\rho_B(c) = \sum c_0 \otimes c_1$ respectively, then their tensor product vector space $B \otimes C$ becomes a (right) $H$-comodule through the linear map $\rho_{B \otimes C} : B \otimes C \to B \otimes C \otimes H$ given by $\rho_{B \otimes C} = (id_B \otimes id_C \otimes m_H) \circ (id \otimes \tau \otimes id) \circ (\rho_B \otimes \rho_C)$. We can straightforwardly check that $\rho_{B \otimes C}$ can be written explicitly: $\rho_{B \otimes C}(b \otimes c) = \sum b_0 \otimes c_0 \otimes b_1 c_1$ establishing thus a (right) $H$-comodule structure for the tensor product of two (right) $H$-comodules.

In the above (and in what follows) we employ the Sweedler's notation for the comodules, according to which $b_i, c_i \in H$ for any $i \neq 0$. We will also use the Sweedler's notation for the comultiplication, according to which $\Delta_C : C \to C \otimes C$ will be written $\Delta_C(c) = \sum c_{(1)} \otimes c_{(2)}$. Finally, we have denoted with $\tau : H \otimes C \to C \otimes H$ the transposition map $\tau(h \otimes c) = c \otimes h$ (which is obviously a v.s. isomorphism). $b, c, h$ are any elements of $B, C, H$ respectively and with $m_H$ we have denoted the multiplication of the Hopf algebra $H$ itself.

Let us now proceed to the main body of the proof:

**Definition A.1:** First of all $C$ being a (right) $H$-comodule coalgebra means that:

**a.** $C$ is a right $H$-comodule (with the coaction denoted by $\rho_C$).

**b.** Its structure maps *i.e.*, the comultiplication $\Delta_C : C \to C \otimes C$ and the counity $\varepsilon_C : C \to \mathbb{C}$, are H-comodule morphisms.

The second statement of the above definition is equivalent (by definition) to the commutativity of the following diagrams

$$
\begin{array}{ccc}
C & \xrightarrow{\Delta_C} & C \otimes C \\
\downarrow \rho_C & & \downarrow \rho_{C \otimes C} \\
C \otimes H & \xrightarrow{\Delta_C \otimes id_H} & C \otimes C \otimes H
\end{array}
\quad \text{and} \quad
\begin{array}{ccc}
C & \xrightarrow{\varepsilon_C} & \mathbb{C} \\
\downarrow \rho_C & & \downarrow \rho_{\mathbb{C}} \\
C \otimes H & \xrightarrow{\varepsilon_C \otimes id_H} & \mathbb{C} \otimes H
\end{array}
\quad \text{(A.1)}
$$

In the above we have made use of the trivial right comodule structure of the field of complex numbers given by $\rho_{\mathbb{C}} : \mathbb{C} \to \mathbb{C} \otimes H$ and explicitly $\rho_{\mathbb{C}}(1) = 1 \otimes 1_H$. The commutativity of the above diagrams is equivalent to the following relations

$$(\rho_{C \otimes C} \circ \Delta_C)(c) = ((\Delta_C \otimes id_H) \circ \rho_C)(c) \Leftrightarrow \sum \sum c_{(1)_0} \otimes c_{(2)_0} \otimes c_{(1)_1} c_{(2)_1} = \sum \sum c_{0_{(1)}} \otimes c_{0_{(2)}} \otimes c_1 \quad \text{(A.2)}$$

$$(\rho_{\mathbb{C}} \circ \varepsilon_C)(c) = ((\varepsilon_C \otimes id_H) \circ \rho_C)(c) \Leftrightarrow \varepsilon_C(c) 1_H = \sum \varepsilon_C(c_0) c_1 \quad \text{(A.3)}$$

Now, if we specialize to the case in which $H = \mathbb{C}G$ *i.e.*, the Hopf algebra itself is the group Hopf algebra then

$$\rho_{B \otimes C}(b \otimes c) = \sum_{g,h \in G} b_g \otimes c_h \otimes gh = \sum_{g,k \in G} b_g \otimes c_{g^{-1}k} \otimes k = \sum_{k \in G} (\sum_{g \in G} b_g \otimes c_{g^{-1}k}) \otimes k \qquad (A.4)$$

But at the same time, we have (by definition)

$$\rho_{B \otimes C}(b \otimes c) = \sum_{k \in G} (b \otimes c)_k \otimes k \qquad (A.5)$$

Equating the coefficients of the rhs of relations (A.4) and (A.5) we get:

$$(b \otimes c)_g = \sum_{h \in G} b_h \otimes c_{h^{-1}g} \qquad (A.6)$$

In the above—and for the sake of clarity—we have slightly digressed from the Sweedler's notation of the coactions, by using the—more explicit—summation notation $\rho_B(b) = \sum_{g \in G} b_g \otimes g$ for the coaction

$\rho_B : B \to B \otimes \mathbb{C}G$ and $\rho_C(c) = \sum c_h \otimes h$ for the coaction $\rho_C : C \to C \otimes \mathbb{C}G$.

Using relation (A.6) in order to re-express the commutativity of the diagrams (A.1) (which is equivalent to the relations (A.2) and (A.3)) we get from (A.2)

$$(\rho_{C \otimes C} \circ \Delta_C)(c) = ((\Delta_C \otimes id_H) \circ \rho_C)(c) \Leftrightarrow \sum \rho_{C \otimes C}(c_{(1)} \otimes c_{(2)}) = \sum_{k \in G} \Delta_C(c_k) \otimes k \Leftrightarrow$$

$$\Leftrightarrow \sum_{k \in G} \sum (\sum_{g \in G} c_{(1)_g} \otimes c_{(2)_{g^{-1}k}}) \otimes k = \sum_{k \in G} \sum c_{k_{(1)}} \otimes c_{k_{(2)}} \otimes k$$

Equating the coefficients of the last relation, with respect to $k \in G$, we finally get

$$\sum (\sum_{g \in G} c_{(1)_g} \otimes c_{(2)_{g^{-1}k}}) = \sum c_{k_{(1)}} \otimes c_{k_{(2)}} = \Delta_C(c_k) \Leftrightarrow \Delta_C(c_k) = \sum (\sum_{g \in G} c_{(1)_g} \otimes c_{(2)_{g^{-1}k}}) \qquad (A.7)$$

Recalling now that since $H = \mathbb{C}G$, the first statement of Definition A.1 is equivalent to the fact that $C = \oplus_{g \in G} C_g$ *i.e.*, $C$ is a $G$-graded vector space, (A.7) implies that

$$\Delta(C_\kappa) \subseteq \oplus_{g \in G} C_g \otimes C_{g^{-1}\kappa} \equiv \oplus_{gh=\kappa} C_g \otimes C_h \qquad (A.8)$$

$$(\rho_{\mathbb{C}} \circ \varepsilon_C)(c) = ((\varepsilon_C \otimes id_H) \circ \rho_C)(c) \Leftrightarrow \rho_{\mathbb{C}}(\varepsilon_C(c)) = (\varepsilon_C \otimes id_H)(\sum c_g \otimes g) \Leftrightarrow$$

$$\Leftrightarrow \varepsilon_C(c) \otimes 1_G = \sum_{g \in G} \varepsilon_C(c_g) \otimes g \Leftrightarrow \varepsilon_C(c)1_G = \sum_{g \in G} \varepsilon_C(c_g)g \Leftrightarrow \begin{cases} \varepsilon_C(c_g) = 0, & \forall g \neq 1_G \\ \varepsilon_C(c_{1_G}) = \varepsilon_C(c), & \forall c \in C \end{cases} \qquad (A.9)$$

Similarly, working out (A.3) produces that

In the last implication, in order to equate the coefficients with respect to $g \in G$, we have used the fact that the elements of the group $G$ are linearly independent (constituting a basis) inside the group algebra $\mathbb{C}G$.

Finally, (A.8) and (A.9) conclude the proof. $\qquad \square$

Let us also note that the above proved implication (and its converse which can be relatively easily filled in) does not depend on $G$ being neither finite nor abelian. In fact the "comodule view", of the grading of a coalgebra $C$ by a group $G$ as being equivalent to a "suitable" coaction (suitable in the sense that the structure maps of the coalgebra $\Delta_C, \varepsilon_C$ become $\mathbb{C}G$-comodule morphisms) of the group

Hopf algebra $\mathbb{C}G$ on $C$, *i.e.*, the equivalence 1.$\Leftrightarrow$3. of Proposition 3.2 is valid for the general case of an arbitrary group. It is the equivalence 2.$\Leftrightarrow$3. of the statements of Proposition 3.2 which is based on $G$ being finite and abelian. For the proof of the later we have to recall that the action of a finite dimensional Hopf algebra on an algebraic structure is equivalent to the coaction of the dual Hopf algebra on the same algebraic structure (and conversely, see [54]) and then to apply the Hopf algebra isomorphism $\mathbb{C}G \cong (\mathbb{C}G)^*$ between $\mathbb{C}G$ and its dual Hopf algebra $(\mathbb{C}G)^*$ which is valid for finite, abelian groups. (We have denoted $(\mathbb{C}G)^* = Hom(\mathbb{C}G, \mathbb{C}) \cong Map(G, \mathbb{C}) = \mathbb{C}^G$ as complex vector spaces and with $\mathbb{C}^G$ we denote the complex vector space of the set-theoretic maps from the finite abelian group $G$ to $\mathbb{C}$).